\documentclass[fleqn,10pt, twocolumn]{wlscirep}
\usepackage[utf8]{inputenc}
\usepackage[T1]{fontenc}

\definecolor{airforceblue}{rgb}{0.36, 0.54, 0.66}
\definecolor{brickred}{rgb}{0.8, 0.25, 0.33}
\definecolor{amber}{rgb}{1.0, 0.75, 0.0}
\definecolor{applegreen}{rgb}{0.55, 0.71, 0.0}
\definecolor{magenta}{rgb}{0.965, 0, 0.859}

\usepackage{nameref}

\title{The gold-rush effect: how innovation speeds up}

\author[1,2,3,*]{Alessandro Bellina}
\author[4,3,2]{Gabriele Di Bona}
\author[1,5,6]{Giordano De Marzo}
\author[1,3,2,6]{Vittorio Loreto}

\affil[1]{Sapienza University, Physics Dept., P.le A. Moro, 2, I-00185 Rome, Italy}
\affil[2]{Centro Studi e Ricerche Enrico Fermi (CREF), Via Panisperna 89/A, 00184 Rome, Italy}
\affil[3]{Sony Computer Science Laboratories - Rome (Sony CSL-Rome), Joint Initiative CREF-SONY, Via Panisperna 89/A, 00184, Rome, Italy}
\affil[4]{CNRS, GEMASS, 59 rue Pouchet, F-75017, Paris, France}
\affil[5]{University of Konstanz, Universitaetstrasse 10, 78457 Konstanz, Germany}
\affil[6]{Complexity Science Hub Vienna, Metternichgasse 8, 1030 Wien}
\affil[*]{Corresponding author. Email: \texttt{alessandro.bellina@cref.it}}

\begin{abstract}

Innovation records often exhibit "hockey-stick" patterns of abrupt, near-singular growth at the collective level. However, this macroscopic explosiveness stands in stark contrast to individual discovery, which remains bounded by cognitive and temporal constraints and follows slow, sublinear accumulation laws. Here, we resolve this micro-macro discrepancy by introducing a minimal multi-scale model that identifies the growth of the explorer population as the primary driver of aggregate acceleration. Building on the Theory of the Adjacent Possible and the Urn Model with Triggering (UMT), we demonstrate that as discoveries expand the space of possibilities, they attract new explorers through a self-reinforcing branching process. This expansion induces a nonlinear mapping between intrinsic time (individual discovery events) and natural time (calendar years), effectively reparameterizing steady individual trajectories into accelerating system-level dynamics. We validate the framework using large-scale patent (EPO) and scientific publication (OpenAlex) datasets, showing that the model accurately reproduces stable per-capita productivity alongside exponential aggregate growth. By providing a quantitative link between individual behavior and collective takeoffs, this work offers a unified foundation for understanding the statistical structure and temporal evolution of innovation ecosystems.
\end{abstract}

\begin{document}

\flushbottom
\maketitle

\thispagestyle{empty}

\section{Introduction}

   In scientific, technological, biological, and socioeconomic systems, empirical records reveal recurring phases of rapid acceleration and near-singular growth~\cite{koppl2023explaining, korotayev202021st, kurzweil2005singularity, von1960doomsday, grinin2015modeling}. Trajectories, from technological performance to global patenting and scientific production, show “hockey-stick” patterns~\cite{cortes2022biocosmology}. These patterns have long periods of slow accumulation, followed by abrupt takeoffs~\cite{korotayev2020twenty, panov2005scaling, nazaretyan2015megahistory}. Recent analyses suggest this rapid expansion can also reshape knowledge circulation and career-level visibility in science~\cite{houssard2024gerontocracy}. The Theory of the Adjacent Possible (TAP) and related growth formalisms~\cite{cortes2025tap, koppl2018simple, steel2020dynamics, bellina2025modelling} formalize these dynamics, predicting super-exponential growth and finite-time singularities. However, these descriptions mainly ignore the microscopic mechanisms of discovery and the role of bounded individual exploration in creating such aggregate accelerations.
    
At the individual level, innovation unfolds differently. Novelty arrives as a sequence of discrete events, each one incrementally expanding an agent’s accessible space. Individual discovery works under hard constraints of cognition, time, and resources~\cite{salmon1994bounded, shaklee1979bounded}. These limits motivate microscopic generative models. Examples include Pólya's urn dynamics~\cite{polya1930quelques, mahmoud2008polya}, Yule–Simon processes~\cite{yule1925ii, simon1955class}, and the Urn Model with Triggering (UMT)~\cite{tria2014dynamics, loreto2016dynamics}. In these models, reinforcement competes with the gradual expansion of possibilities. Such models account for slow novelty production, including sublinear Heaps’ laws~\cite{herdan1960type, heaps1978information}, heavy-tailed frequency distributions~\cite{newman2005power}, and diminishing returns~\cite{boldrin2009model}. Their trajectories use an intrinsic clock—accumulated discovery events rather than calendar time—so they do not generate the macroscopic accelerations observed historically. While crucial for explaining Heaps’ and Zipf’s laws~\cite{zipf1929relative, zipf2013psycho, li2002zipf}, microscopic triggering mechanisms alone cannot explain the explosiveness of large-scale innovation.

A formal bridge between micro and macro perspectives is still missing. Existing approaches capture either slow, bounded individual discovery or rapid, accelerating aggregate patterns, but not both within a single framework. This raises questions: How can sublinear individual paths generate aggregate superlinear or hyperbolic patterns? How do many independent exploratory paths produce collective accelerations? Which mechanism maps individual intrinsic time to natural calendar time?

This micro–macro discrepancy is clear in empirical records. In patent and publication data, the interval between outputs for an inventor or author often remains stable over a career~\cite{huber2001new, cole1979age}. By contrast, at the system level, events occur closer together, showing collective acceleration~\cite{de1963little, marco2015uspto, beggs1984long}. Individual output usually grows linearly over a career, with production rates stable or gradually declining~\cite{fanelli2016researchers, larsen2010rate}. Yet, total innovation grows superlinearly and sometimes approaches hyperbolic scaling~\cite{youn2015invention, fortunato2018science, nazaretyan2020twenty}. Meanwhile, the population of active explorers in science and technology rises sharply over the course of decades~\cite{prichina2017world, gascoigne1992historical, zapp2022revisiting, larsen2010rate}. Altogether, these patterns show that aggregate acceleration results from collective-level dynamics, not faster individual progress.
    
A likely mechanism is the fast growth of the explorer population, which co-evolves with the expanding space of possibilities. This can generate “gold-rush” entry dynamics~\cite{dosi1982technological, lee1966theory, taylor1999new}. When more explorers enter, the total number of discovery events per year increases—even if each individual works at a steady pace~\cite{larsen2010rate, fanelli2016researchers}. Population growth creates a nonlinear mapping between intrinsic time (individual discoveries) and natural time (years). This mapping can turn slow individual progress into linear, superlinear, or faster macroscopic growth. Thus, aggregate acceleration is an amplification of many parallel explorations.
    
Guided by this picture, we introduce a minimal unifying model that links microscopic discovery dynamics to macroscopic innovation by treating the growth of the explorer population as an explicit state variable. Grounded in the Theory of the Adjacent Possible~\cite{kauffman1992origins, tria2014dynamics},
%~\cite{packard1988adaptation, langton1990computation, kauffman1992origins, tria2014dynamics}
the framework extends the Urn Model with Triggering (UMT)~\cite{tria2014dynamics, loreto2016dynamics, bellina2025full} to a multi-agent setting. Multiple explorers jointly enlarge the space of potential novelties. The explorer entry follows a branching process that models how new opportunities attract new participants. The resulting dynamics highlight the minimal ingredients needed to translate bounded individual exploration into accelerating collective change.

The model shows that collective takeoffs can arise even if individual discovery stays stable. As the explorer population grows, many slow trajectories add up in parallel. Aggregate growth in calendar time can become superlinear, exponential, or nearly hyperbolic, depending on the system. This gives a concrete microscopic reason why innovation records can accelerate even without faster individual effort.
    
\section{Results}
\label{sec:results}

    \subsection{From microscopic individual discovery to macroscopic collective innovation}
    \label{sec:results_1} 
    
    \begin{figure*}[ht!]
        \centering
        \includegraphics[width=\linewidth]{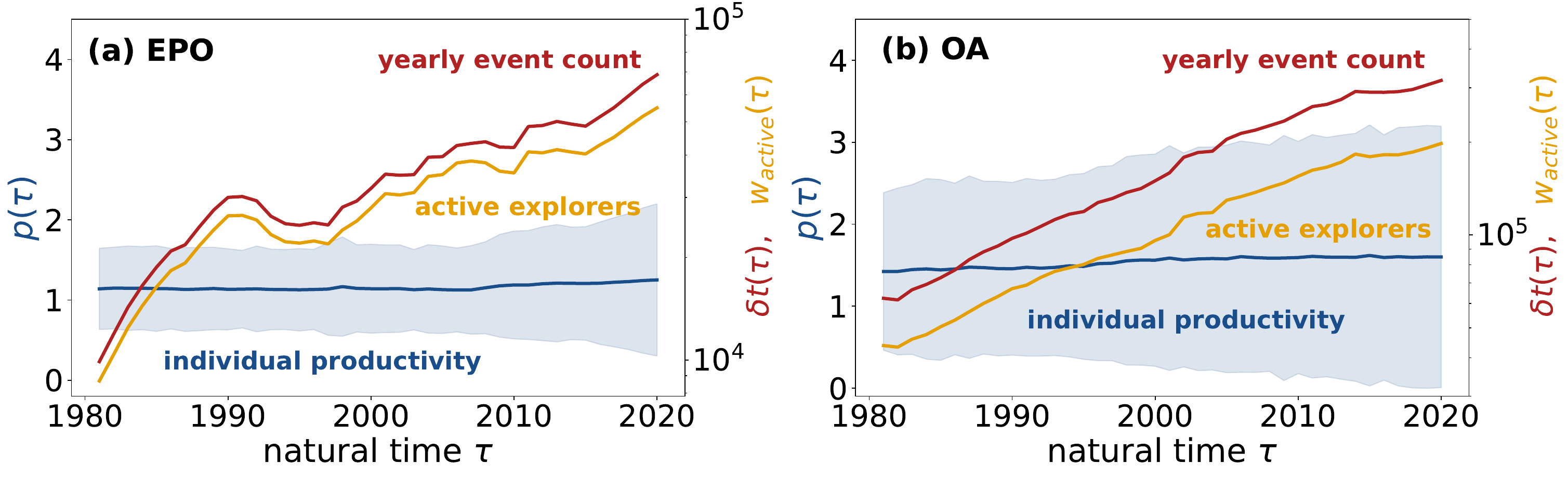}
        \caption{
        \textbf{Collective acceleration vs. individual stability in scientific and technological production.} 
        We track yearly event count, $\delta t(\tau)$ (events per year), number of active explorers, $w_{\mathrm{active}}(\tau)$, and per-capita productivity $p(\tau)=\delta t(\tau)/w_{\mathrm{active}}(\tau)$ for patents from the European Patent Office (EPO) \textbf{(a)} and scientific publications from the OpenAlex (OA) database \textbf{(b)}, where natural time $\tau$ is measured in years. Each panel presents per-capita productivity (blue curve) on the left vertical axis (linear scale), and yearly event count (red curve) together with active explorers (yellow curve) on the right vertical axis (logarithmic scale). In both domains, $\delta t(\tau)$ and $w_{\mathrm{active}}(\tau)$ grow approximately exponentially, indicating strong collective acceleration. In contrast, per-capita productivity remains essentially constant. Thus, macro-level acceleration arises from rapid growth in the explorer population, not from faster individual discovery, demonstrating that distinct temporal scales govern discovery and innovation.
        }
        \label{fig:fig1}
    \end{figure*}
    
The data show a clear contrast between innovation at the individual and collective levels. Over decades, the total number of papers and patents produced accelerates, often approaching exponential growth in calendar time~\cite{bellina2025modelling, koppl2023explaining, korotayev202021st}. Meanwhile, individual exploration occurs through discrete discovery events and, when tracked in intrinsic time (i.e., simply counting such events), follows sublinear laws of novelty accumulation~\cite {heaps1978information, tria2014dynamics, di2025dynamics}. Career-level studies find the average output rate of a typical author or inventor remains stable throughout a career~\cite{huber2001new, larsen2010rate}. How do bounded individual dynamics coexist with, and ultimately generate, accelerating aggregate growth?

This contrast between individual and collective innovation is clear in large-scale scientific and technological records. We examine two examples: the EPO patent repository~\cite{european1991european} and the \mbox{OpenAlex} corpus of scientific publications (OA)~\cite{priem2022openalex}. Figure~\ref{fig:fig1}a--b shows the yearly number of patents and publications produced, $\delta t(\tau)$. In both, $\delta t(\tau)$ rises rapidly over time and fits exponential growth, showing strong collective acceleration.

A similar pattern appears for the number of active explorers, $w_{\mathrm{active}}(\tau)$. This denotes authors or inventors who contribute at least one item in a given year. As shown in Fig.~\ref{fig:fig1}a--b, this population rises sharply over time, tracking system-level output. The joint rise of $\delta t(\tau)$ and $w_{\mathrm{active}}(\tau)$ fits the rapid takeoffs in adjacent-possible formalisms like the TAP equations. These models describe accelerating growth across biological, technological, and socioeconomic systems~\cite{cortes2025tap, koppl2023explaining, korotayev202021st}.
    
Despite these rapid aggregate increases, individual productivity $p(\tau)$ remains remarkably stable over time (Fig.~\ref{fig:fig1}). To see this, consider that when normalized by the number of active contributors, the average output per explorer shows no evidence of acceleration. This finding is consistent with approximately stationary inter-event times at the individual level and with cumulative output that is typically linear or mildly sublinear in career time. On average, an active explorer produces $\mathcal{O}(1)$ patents or publications per year. The full distribution of yearly individual productivity is reported in the Supplementary Information (Section~\ref{sec:SI2}), where explorers display substantial heterogeneity around a stable average behavior.
    
This apparent stationarity merits closer examination in light of previous studies reporting mildly superlinear productivity at the individual level~\cite{petersen2012persistence,sinatra2016quantifying}. Those studies condition on career time by tracking authors from their first publication. In contrast, our approach averages over all authors active in a given calendar year—who are simultaneously at very different career stages—so life-cycle effects are naturally smoothed out. This difference in approach reveals a clear separation of scales: slow, bounded individual trajectories coexist with rapid system-level expansion. Macroscopic acceleration arises not from faster individual exploration, but from the growing number of explorers active in the system.
    
This separation between individual stability and system-level acceleration is echoed at the modeling level. Triggering-based models of discovery, such as the Urn Model with Triggering (UMT)~\cite{tria2014dynamics,loreto2016dynamics, bellina2025full}, describe exploration as unfolding in an intrinsic time variable $t$, counting individual discovery events. In intrinsic time, novelty accumulation follows Heaps-like laws, and productivity profiles remain stable. However, intrinsic time captures individual dynamics rather than aggregate activity, so such models alone do not reproduce the rapid surges observed in calendar time. The key empirical observation connecting model and data is that the growth of the explorer population closely tracks the growth of system-level output (Fig.~\ref{fig:fig1}): as $w_{\mathrm{active}}(\tau)$ increases, more discovery events occur per unit calendar time even when per-capita productivity remains approximately constant. This observation points to population expansion as the missing link between steady intrinsic-time exploration and accelerating aggregate dynamics.

In summary, these observations indicate that the missing link between microscopic discovery and macroscopic acceleration is the growth of the active explorer population. This mechanism induces a nonlinear mapping between intrinsic time (event counts) and natural time (years). In the next section, we formalize this mechanism with a minimal multi-scale model.

\subsection{A multi-scale model of innovation dynamics}
\label{sec:results_2}
    
\begin{figure*}[htb!]
    \centering
    \includegraphics[width=\linewidth, trim=6cm 1cm 6cm 1cm, clip]{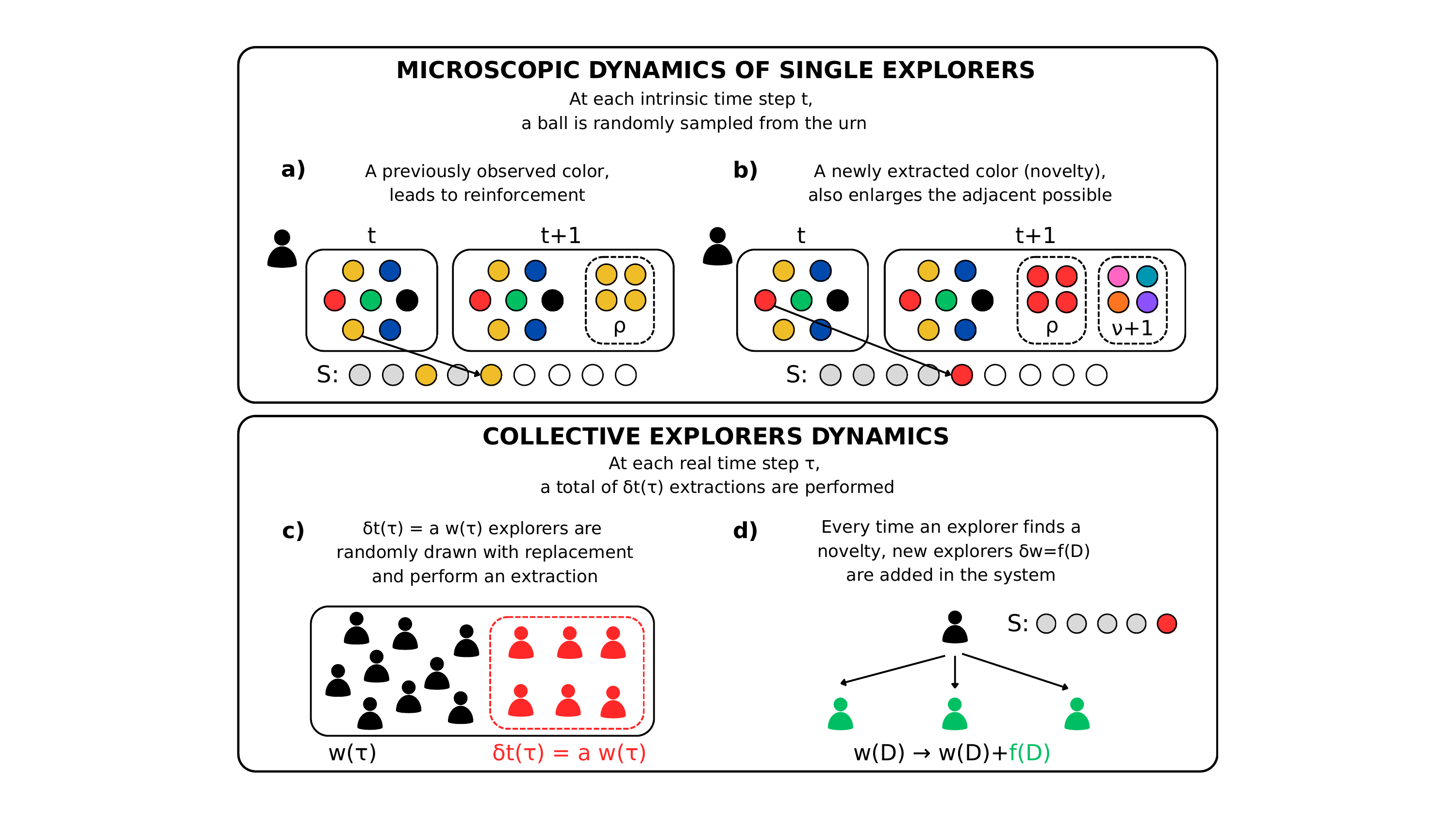}
    \caption{\textbf{Schematic illustration of the microscopic and collective components of the model.} \textbf{(a--b) Microscopic UMT dynamics.} At each intrinsic-time step, an explorer extracts an element from the urn. \textbf{(a)} If the color has appeared before, it is reinforced by returning it with $\rho$ copies. \textbf{(b)} If it is new, the adjacent possible expands: $\rho$ reinforcement copies and $\nu+1$ brand-new colors are added to the urn. The extracted element is appended to the collective \emph{exploration} sequence $S$. \textbf{(c--d) Collective dynamics.} \textbf{(c)} Each novelty triggers the arrival of additional explorers following the branching rule $\delta w = f(D)$. \textbf{(d)} In natural time $\tau$, events occur at a rate proportional to the number of active explorers, $\delta t(\tau) = a\, w(\tau)$; at each step, an explorer is sampled uniformly with replacement and follows the UMT rules. Together, these mechanisms couple microscopic exploration with population growth, providing the minimal ingredients for macroscopic innovation accelerations.}
\label{fig:fig2}
\end{figure*}
    
We now introduce a minimal multi-scale model that couples microscopic adjacent-possible exploration to endogenous population growth, thereby linking intrinsic-time discovery dynamics to accelerating trends in calendar time (Fig.~\ref{fig:fig2}). The model comprises three components: (i) a microscopic discovery process that governs how novelties arise along individual trajectories; (ii) a branching mechanism, in which novelties attract additional explorers; and (iii) a productivity relation that maps intrinsic time to natural time by setting the event rate proportional to the active population.

Intrinsic time $t$ counts the cumulative number of exploration events (one extraction per event), aggregated across all active explorers, while natural time $\tau$ denotes calendar time (years or days). The explorer population, $D_w(t)$, represents the number of agents present on the network at time $t$, namely those who have produced at least one discovery event. In the baseline model, we simplify by neglecting exit: explorers introduced via novelty production remain active indefinitely (a possible extension including mortality is discussed in the Supplementary Information, Section~\ref{sec:SI1}). Thus, the number of active explorers and the cumulative number of explorers are equivalent. In natural time, the empirical proxy $w_{\mathrm{active}}(\tau)$ captures contributors producing at least one item in a given year, and $D_w(\tau)$ tracks the cumulative number of distinct explorers observed up to time $\tau$.

\begin{figure*}[htb!]
\centering
    \includegraphics[width=\linewidth]{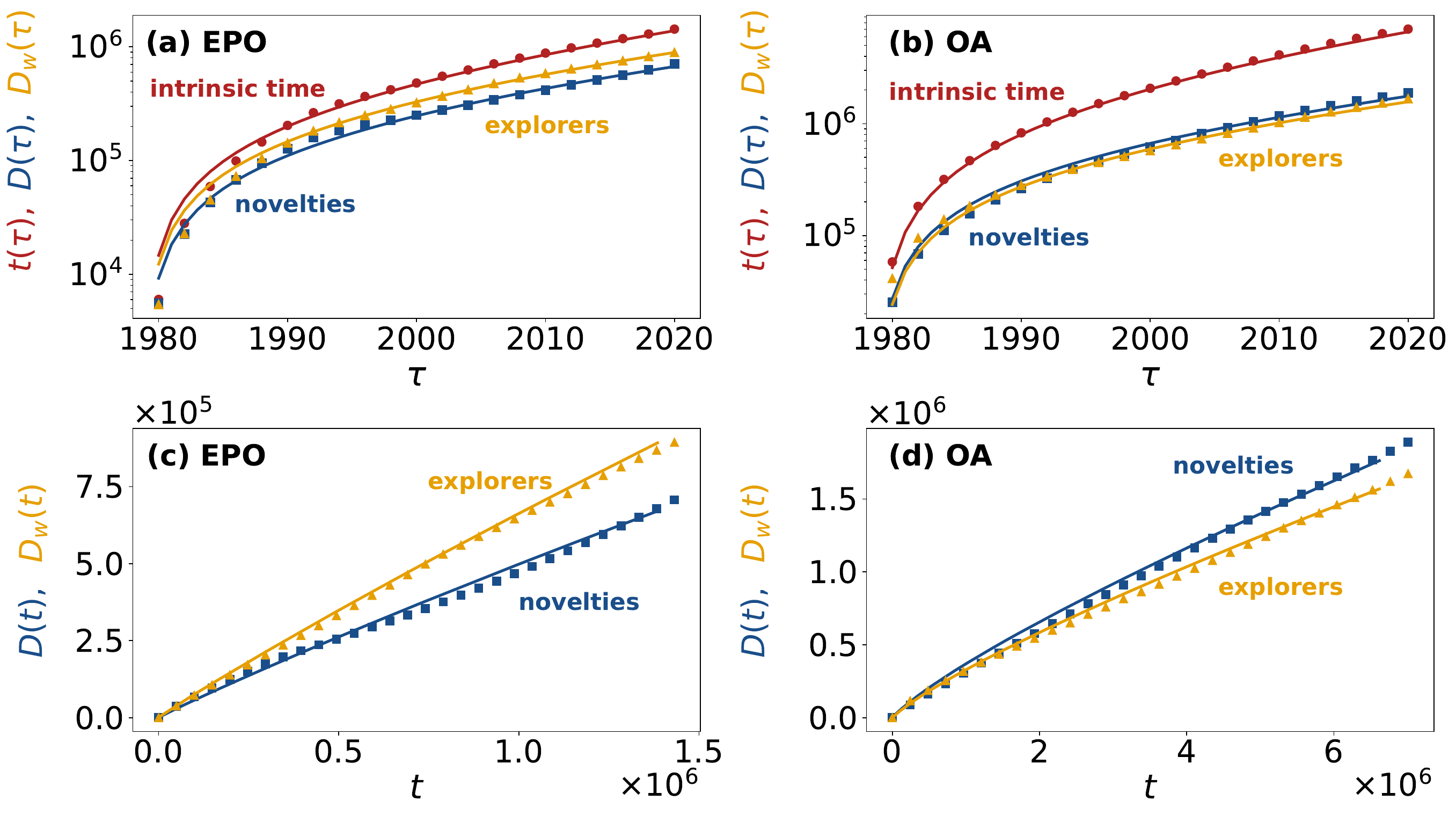}
    \caption{
        \textbf{Model--data comparison in natural and intrinsic time.} \textbf{Top panels (a--b):} empirical data (markers) and model simulations (solid lines) of the cumulative number of events $t(\tau)$, the number of novelties $D(\tau)$, and the cumulative number of distinct explorers $D_w(\tau)$ as functions of natural time $\tau$ for patents (EPO) \textbf{(a)} and scientific publications (OA) \textbf{(b)}. Here $t(\tau)$ counts all patents or publications produced, $D(\tau)$ counts novelties defined as first appearances of new technological or scientific \emph{combinations} (IPC-code combinations or keyword combinations), and $D_w(\tau)$ counts the cumulative number of distinct inventors/authors observed up to time $\tau$. Calibrated through the branching and productivity relations, the model reproduces the empirical growth of all three quantities and captures the acceleration induced by the expanding explorer population. \textbf{Bottom panels (c--d):} the same quantities expressed in intrinsic time $t$, which counts cumulative discovery events. In this representation, both novelties and explorer introductions follow approximately linear microscopic trends, showing that the coupled urn--branching dynamics provides a consistent description across temporal scales.
        }
\label{fig:fig3}
\end{figure*}

At the microscopic level, we use the Urn Model with Triggering (UMT)~\cite{tria2014dynamics}. This model captures reinforcement and adjacent-possible expansion in an analytically tractable way. For an individual explorer, the probability of producing a novelty determines how the number of unique discoveries $D(t)$ evolves in intrinsic time $t$. In mean-field form, novelty accumulation grows sublinearly (or at most linearly):
\begin{equation*}
    \frac{dD}{dt} \approx \frac{\nu\, D(t)}{\rho\, t + (\nu + 1)\, D(t)}.
\end{equation*}
Here, $D(t)$ counts the number of \emph{new} combinations encountered, such as first-time IPC or keyword pairs. $t$ is the total number of events, including repetitions. In the regime $\nu=\rho=1$, which we adopt here, novelty accumulation is nearly linear in intrinsic time (with logarithmic corrections). It is well approximated as linear over the empirical range: $D(t)\approx t$.
    
To capture the empirical tendency of high-innovation domains to attract more contributors, we assume that the explorer population increases only when new novelties appear. That is, the model explicitly assumes that whenever a novelty is produced, a fixed proportion of new explorers is introduced, establishing a linear branching relationship between novelty production and population growth:
\begin{equation}
        \frac{dw}{dD} = b
        \qquad \Rightarrow \qquad
        w(D) = w_0 + b\, D \equiv w_0 + D_w,
        \label{eq:branching}
\end{equation}
where $w_0$ is the initial number of explorers, and $D_w=bD$ represents the explorers introduced through novelty production. We discuss more general branching processes in Section~\ref{sec:methods_model}.
    
To relate intrinsic time $t$ to natural time $\tau$, we further assume that each time increment in calendar time arises from exploration events occurring at a rate proportional to the active explorer population. This assumption links the timing of aggregate events to current system activity.
\begin{equation}
        \frac{dt}{d\tau} = a\, w\!\bigl(t(\tau)\bigr) = a\,[w_0 + D_w(\tau)],
        \label{eq:productivity}
\end{equation}
where $a$ is an effective per-explorer contribution. In our implementation, at each step, an explorer is sampled uniformly from all active ones up to a vanishing correction induced by branching; the parameter $a$ therefore reflects the average individual productivity.
    
By combining the productivity relation, microscopic UMT dynamics, and linear branching, we obtain in natural time:
\[
        \frac{dD}{d\tau}
        = \frac{dD}{dt}\,\frac{dt}{d\tau}
        = a\, w\!\bigl(t(\tau)\bigr)\,\frac{dD}{dt}
        = a\,[w_0 + b\,D(\tau)]\,\frac{dD}{dt}.
\]
In the regime $\nu=\rho=1$ adopted here, novelty accumulation is near-linear in intrinsic time over the empirical range (up to logarithmic corrections), so that $dD/dt \approx 1$. Under this approximation,
\[
        \frac{dD}{d\tau} \simeq a\,[w_0 + b\,D(\tau)],
\]
which integrates to exponential growth in natural time even though $D(t)$ is only (near-)linear in intrinsic time. Under the approximation $dD/dt \approx 1$ and with the initial condition $D(\tau{=}0)=0$ (start of the observation window), integration yields
\begin{equation*}
        D(\tau) \simeq \frac{w_0}{b}\,\bigl(e^{ab\tau}-1\bigr).
\end{equation*}
Using $D_w(\tau)=bD(\tau)$ gives
\begin{equation*}
        D_w(\tau) \simeq w_0\,\bigl(e^{ab\tau}-1\bigr),
\end{equation*}
and integrating Eq.~\eqref{eq:productivity} gives the cumulative number of events up to time $\tau$,
\begin{equation*}
        t(\tau) \simeq \frac{w_0}{b}\,\bigl(e^{ab\tau}-1\bigr),
\end{equation*}
which satisfies $t(0)=0$. Therefore, in natural time, the three quantities $D(\tau)$, $D_w(\tau)$, and $t(\tau)$ all grow exponentially with characteristic rate $ab$, while remaining bounded (linear) in intrinsic time. Figure~\ref{fig:fig2} summarizes how microscopic UMT dynamics, branching-driven population growth, and the productivity mapping jointly generate the multi-scale behavior captured by the model.
    
The microscopic urn dynamics also determines the frequency distribution of explored elements, yielding a power-law form $p(f)\sim f^{-1-\nu/\rho}$, which depends explicitly on the urn parameters~\cite{tria2014dynamics, loreto2016dynamics}. Because this distribution emerges directly from individual-level exploration, the model provides access to both aggregate quantities and the detailed statistical structure of the underlying elements. This enables direct comparison with empirical frequency statistics in the next section.

\subsection{Modeling innovation across multiple scales}
    \label{sec:results_3}
    
\begin{figure*}[htb!]
    \centering
    \includegraphics[width=\linewidth]{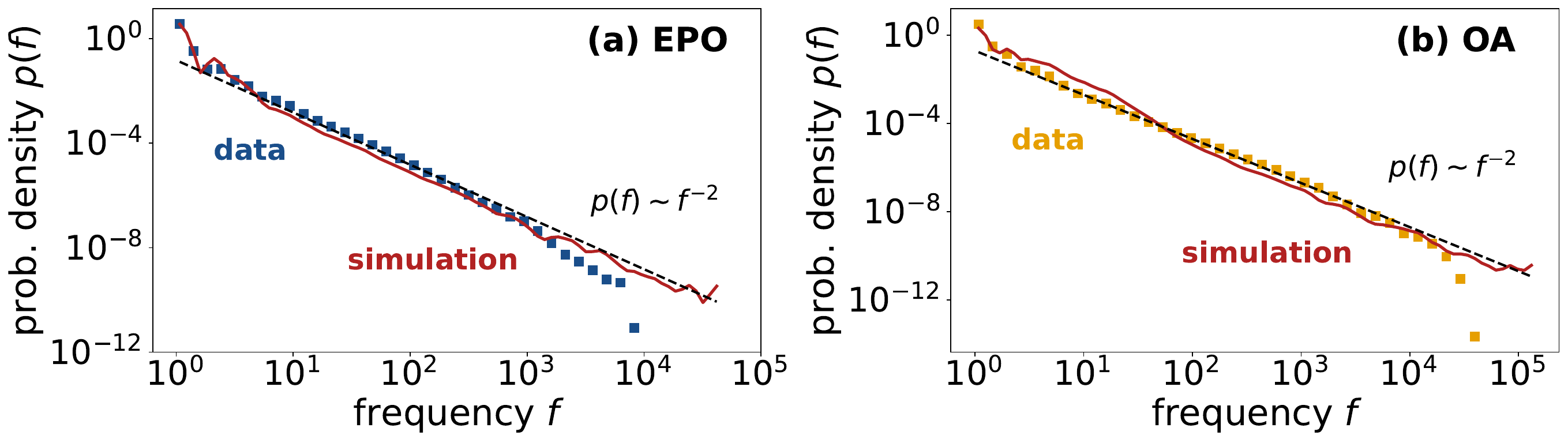}
    \caption{
    \textbf{Frequency distribution of technological and scientific elements.} 
    Empirical data (markers) and model simulations (solid red lines) for patents (EPO) \textbf{(a)} and scientific publications (OA) \textbf{(b)}. The dashed black line indicates the reference scaling $p(f)\sim f^{-2}$. The horizontal axis reports the frequency $f$ of each distinct element (an IPC combination or a scientific keyword), and the vertical axis shows the corresponding probability density $p(f)$ on log--log scales. Both domains display Zipf-like statistics: the probability density follows approximately $p(f)\sim f^{-2}$, corresponding to a Zipf rank--frequency exponent close to one, a common signature of discovery and innovation processes. The model reproduces this heavy-tailed distribution when run with the microscopic urn dynamics in the regime $\nu=\rho=1$, showing that the same minimal mechanism that matches temporal trends also captures the statistical structure of the explored elements.
    }
\label{fig:fig4}
\end{figure*}
    
We now empirically evaluate the framework developed earlier. Despite its minimal structure and few parameters, the model gives a simple and informative view of innovation across scales. At the microscopic level, we use urn dynamics in the parsimonious regime with $\nu=\rho=1$, which is both empirically motivated and sets the frequency statistics for the explored elements. Discovery and innovation often show Zipf-like regularities, typically as a rank–frequency law with exponent near one~\cite{zipf2013psycho, li2002zipf, de2021dynamical}. In frequency space, this means a heavy-tailed distribution with density $p(f)\sim f^{-2}$, which our urn dynamics replicate in this regime. With these settings, the microscopic component is fully defined.
    
At the macroscopic level, we calibrate the model’s two main relations. The first is the branching mechanism, which links novelties with the growth of the explorer population (Eq.~\ref{eq:branching}). The second is the productivity relation connecting intrinsic and natural time (Eq.~\ref{eq:productivity}). The branching relation predicts a linear dependence $D_w = b\, D$, allowing $b$ to be estimated from a simple fit between the number of explorers and novelties. As explained in Section~\ref{sec:methods_calibration}, we obtain $b = 1.33$ for patents and $b = 0.89$ for publications, both with excellent fit.
    
We calibrate the productivity relation by fitting empirical yearly event counts $\delta t(\tau)$ against the number of observed explorers $D_w(\tau)$. In our model, $\delta t(\tau)$ is the discrete-time equivalent of the continuous rate $dt/d\tau$ in Eq.~\ref{eq:productivity}. This fit provides the parameters $a$ and the initial value $w_0$, which is the number of explorers at the start of the observation window. Figure~\ref{fig:fig6} shows the fitting process and the estimated values of $a$ and $w_0$ for both datasets. For both Eq.~\ref{eq:branching} and Eq.~\ref{eq:productivity}, the linearity assumption holds up well and describes the empirical relations accurately.
    
We also account for the initial number of distinct elements, $N_0$, in the urn. This is important when data collection begins after the innovation process has already started. In our datasets, the observations begin in 1980. By then, both fields had already accumulated substantial knowledge. To capture this transient phase, when early-year dynamics is inflated by existing items, we specify the urn’s initial state. We set $N_0 = 1.2\times 10^{5}$ for patents and $N_0 = 1.3\times 10^{4}$ for publications. Section~\ref{sec:methods_initial_cond} discusses this choice and its effects in detail.
    
Using the estimated parameters, we simulate the model for both patents and publications. We generate the same number of events as observed empirically ($t \approx 1.4\times 10^{6}$ for patents, $t \approx 7.0\times 10^{6}$ for publications). Figure~\ref{fig:fig3} shows the results. The top panels (a–b) compare empirical and simulated trajectories for total events $t(\tau)$, novelties $D(\tau)$, and distinct authors $D_w(\tau)$ versus natural time $\tau$. Here, $t(\tau)$ is the total number of patents or publications. $D(\tau)$ counts novelties as the first appearance of new technological or scientific combinations (IPC or keyword combinations). The framework strongly matches the data: it captures the mapping between intrinsic and natural time, the growth of novelty, and the explorer population’s expansion through branching.
    
Panels (c–d) of Fig.~\ref{fig:fig3} present the same quantities in intrinsic time $t$. In this view, the model closely matches empirical behavior. Accumulation of novelties and the growth of author numbers follow predicted trends. This shows the model’s microscopic and macroscopic parts describe the data consistently across time scales.
    
The intrinsic-time view allows detailed comparison of the elements’ statistical properties. The urn dynamics naturally generate frequency distributions for elements (how often each IPC or keyword appears). This allows direct comparison with empirical frequency statistics. Figure~\ref{fig:fig4} shows that the data have a Zipf-like distribution $p(f)\sim f^{-2}$, a common pattern in innovation and discovery. The model reproduces this distribution at parameters $\nu=\rho=1$, providing an accurate microscopic account of appearance frequencies.
    
We do not explicitly model differences in individual productivity. Explorers are sampled uniformly from the active pool, making author event distributions trivial by design. More detail—such as productivity variation and entry or exit—can be added with explicit activity mechanisms, as outlined in the Supplementary Information. The mean behavior matches the data: per-capita productivity $p(\tau)=\delta t(\tau)/w_{\mathrm{active}}(\tau)$ remains $\mathcal{O}(1)$ over the observation window (see Fig.~\ref{fig:fig1}).
    
Overall, with minimal and empirically grounded assumptions, the model matches many empirical patterns in both patents and scientific publications. Using only a few calibrated parameters, it gives a coherent picture of innovation at different scales. The model links collective acceleration in natural time to the limited microscopic behavior of individual explorers.
    
\section{Discussion}
\label{sec:discussion}

Our work addresses a fundamental multiscale tension in innovation. Aggregate indicators such as scientific production and patenting accelerate over historical time. In contrast, individual discovery processes remain slow and bounded. These observations have often been treated with separate modeling approaches operating at different levels. Here, we provide a unified mechanism that links the two. Stable microscopic exploration, when aggregated across a growing population of explorers, can naturally yield superlinear and even near-hyperbolic macroscopic growth. In this view, collective amplification driven by population expansion—rather than a systematic speed-up of individuals—reconciles micro-level regularities with macro-level accelerations. This is consistent with the observed stability of per-capita productivity.

A central implication of the model is the coexistence of two temporal descriptions. The first is an intrinsic exploration time $t$, which counts discovery events and captures microscopic dynamics along individual trajectories. This level displays stable productivity and sublinear (or at most linear) novelty accumulation. The second is the natural time $\tau$ of the system, measured in years, in which many exploratory processes unfold in parallel. Population growth couples these layers by inducing a nonlinear mapping between $t$ and $\tau$. This process effectively reparameterizes steady individual dynamics into accelerating collective trajectories.
    
At its simplest, the model isolates the time-mapping mechanism that converts bounded exploration into collective acceleration. A natural next step is to relax the assumption of homogeneous explorers. Real innovation systems exhibit strong heterogeneity in career trajectories. These are shaped by broad productivity distributions, cumulative advantage~\cite{price1976general, barabasi1999emergence}, and field-specific constraints~\cite{ernst2000inventors, sabharwal2013comparing, allison1974productivity}. Introducing heterogeneous exploration rates or individual triggering parameters would allow the framework to capture second-order statistics—such as distributions of career outputs and burstiness—without altering the core coupling between intrinsic and natural time. We document these heterogeneities empirically and outline corresponding model extensions in the Supplementary Information.
    
Second, the baseline model does not include explicit interactions among explorers. Agents contribute independently to expanding the adjacent possible. However, interaction-driven discovery on networks can strongly modulate exploration dynamics~\cite{iacopini2020interacting}. Collaboration is a central driver of scientific and technological production. Co-authorship networks, knowledge diffusion mechanisms, and team structures introduce dependencies among exploratory paths and can shape both the rate and the direction of innovation~\cite{sonnenwald2007scientific, wuchty2007increasing, sinatra2016quantifying, di2022social}. Incorporating interaction patterns into the multi-explorer setting would allow the model to capture how network effects modulate microscopic exploration and propagate into macroscopic acceleration. This would link the present mechanism to empirical findings on team-driven creativity.
    
Third, the framework can be extended to structured possibility spaces. In the present model, discovery unfolds in an abstract urn, effectively treating the space of combinations as unstructured. However, empirical evidence indicates that innovation operates on structured domains—semantic networks~\cite{steyvers2005large}, knowledge graphs~\cite{hogan2021knowledge}, and technological spaces~\cite{ivanov2002technological}. The topology of these domains constrains and channels exploration~\cite{di2025dynamics, iacopini2018network, tria2014dynamics}. Embedding such networked spaces would enable systematic investigation of how connectivity, modularity, and local geometry shape the emergence of novelties. It would also help analyze how external perturbations, such as funding shifts, policy interventions, or technological shocks, deform the space and redirect innovation trajectories. For instance, concentrated investment in specific areas, such as artificial intelligence, may locally increase explorer density and induce asymmetric expansions of the adjacent possible.
    
Overall, our results support a multiscale view of innovation. Macroscopic accelerations emerge from the aggregation of many bounded discovery processes under population growth and adjacent-possible expansion. This perspective provides a quantitative bridge for analyzing how interventions at one level can propagate across the system. For example, interventions may include supporting early-career researchers, reshaping collaboration structures, or fostering interdisciplinarity. More broadly, clarifying how individual behavior, population dynamics, and the structure of the possibility space co-determine innovation provides a foundation for predictive and generative modeling. Ultimately, this approach helps answer normative questions about how to steer innovation ecosystems.

\section{Methods}
\label{sec:methods}

\subsection{Data sets}
\label{sec:methods_data}

\paragraph{Technological innovation (EPO / PATSTAT).}
We use patent records from the European Patent Office (EPO) \cite{european1991european}, obtained via the PATSTAT database \cite{jacob2013patstat,de2014introduction}. We analyze all patents filed between 1980 and 2020 ($1{,}432{,}649$ documents), extracting IPC/CPC technological classes, inventors, and filing dates. Each patent is represented by the set of IPC/CPC classification codes assigned to it; we define a novelty as the first occurrence of a previously unseen code \emph{combination} (set). The first listed inventor is used to track the entry of new explorers. The dataset includes $894{,}634$ distinct first inventors and $71{,}782$ IPC/CPC codes at the considered level of granularity. To mitigate extreme productivity outliers and potential disambiguation artifacts, we exclude inventors producing more than $50$ patents in a single year, resulting in the removal of $7$ inventors and $10{,}379$ patents (approximately $0.7\%$ of the dataset).
        
\paragraph{Scientific innovation (OpenAlex).}
Scientific publications are obtained from OpenAlex~\cite{priem2022openalex}. We select works in Mathematics and Physics \& Astronomy published between 1980 and 2020, for a total of $7{,}047{,}964$ articles. Each paper is represented by the set of associated keywords; we define a novelty as the first occurrence of a previously unseen keyword \emph{combination} (set) in the corpus. The first author is used to track the growth of the explorer population. The dataset comprises $1{,}673{,}411$ distinct authors and $35{,}026$ unique keywords. To mitigate disambiguation artifacts and extreme hyper-authorship effects, we exclude authors who produce more than $100$ papers in a single year, removing $58$ authors and $40{,}465$ papers (approximately $0.6\%$ of the dataset).

\begin{figure*}[htb!]
    \centering
    \includegraphics[width=\linewidth]{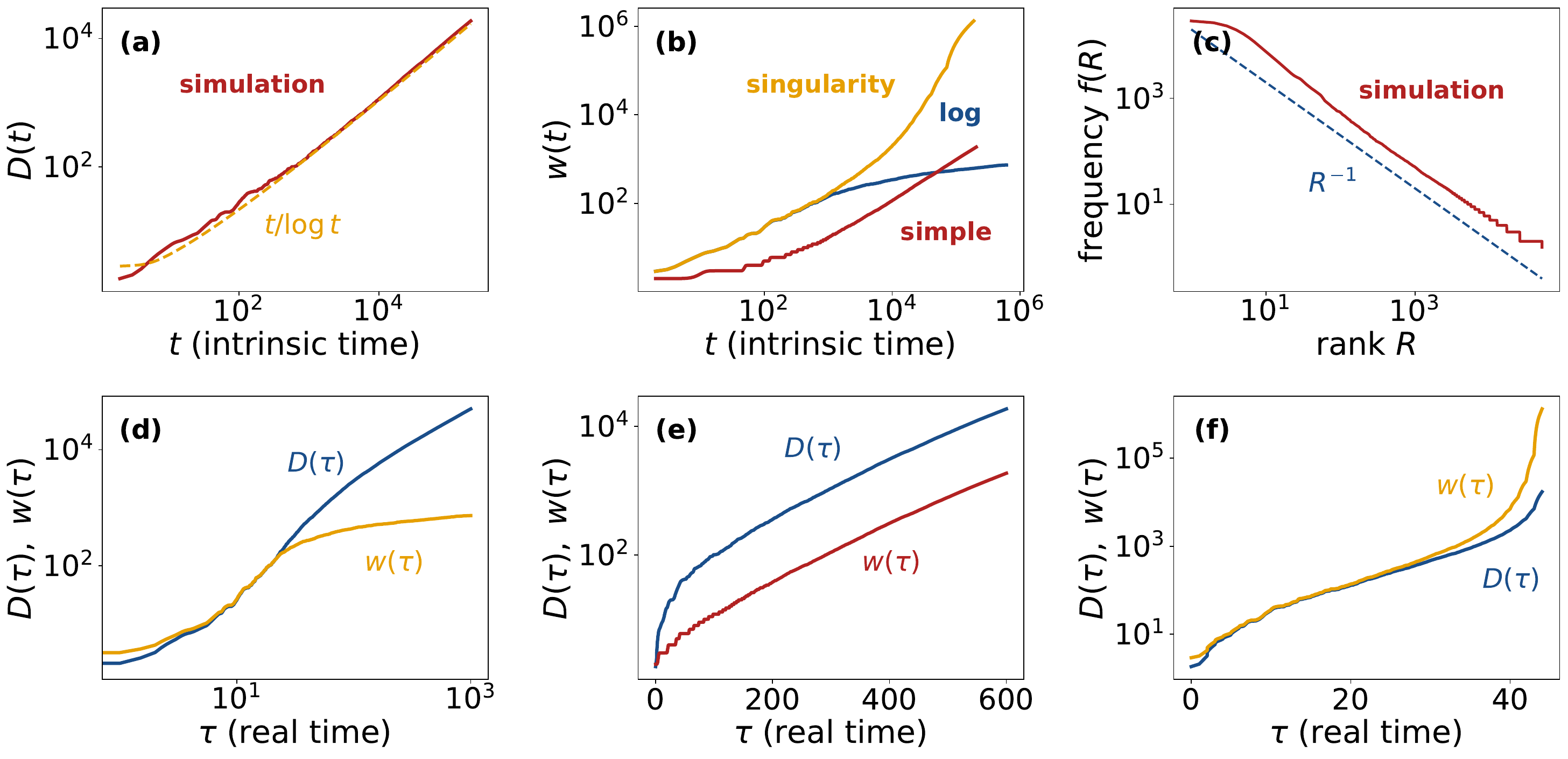}
    \caption{
    \textbf{Model behavior across branching regimes in intrinsic and natural time.} The model is simulated with urn parameters $\nu=\rho=1$, initial condition $N_0=1$, and $a=1$, $w_0=0$. Panels\textbf{(a–c)} show quantities in intrinsic time $t$, while panels \textbf{(d–f)} display the corresponding dynamics in natural time $\tau$, obtained from $dt/d\tau = a w(t(\tau))$. \textbf{(a)} The number of distinct elements follows the UMT prediction $D(t)\sim t/\log t$, identical across branching regimes. \textbf{(b)} Growth of the explorer population $w(t)$ for logarithmic ($w\sim \log t$), linear ($w\sim t$), and exponential ($w\sim e^{t}$) branching processes. \textbf{(c)} Frequency–rank distributions on log--log scales, displaying the Zipf-like behavior $f(R)\sim R^{-1}$ generated by the microscopic urn dynamics. \textbf{(d–f)} Dynamics in natural time $\tau$ for the three branching cases. \textbf{(d)}: Logarithmic branching leads to slow, near–intrinsic behavior; \textbf{(e)}: linear branching produces exponential growth; \textbf{(f)}: exponential branching generates a finite-time singularity, with $D(\tau)$ and $w(\tau)$ diverging super-exponentially. These regimes illustrate how different branching forms lead to distinct macroscopic innovation patterns despite identical microscopic exploration patterns.}
\label{fig:fig5}
\end{figure*}

\subsection{Model}
\label{sec:methods_model}

We model innovation as the outcome of many concurrent explorers navigating an expanding adjacent possible. Each explorer performs elementary discovery events, while each novelty triggers the appearance of further potential elements according to the Urn Model with Triggering (UMT)~\cite{tria2014dynamics}. Let $D$ denote the number of distinct elements discovered, $t$ the intrinsic time (number of extractions), and $\tau$ the natural time (calendar time). Explorers enter the system through a branching process with population size $w$, which grows with novelty production and is observed as a function of natural time $\tau$. 

In the practical implementation of the model, each novelty event triggers the introduction of $f(D)$ new explorers. These explorers are activated immediately by producing their first event in the subsequent intrinsic steps, thereby entering the observed population. As a consequence, branching induces a finite number of intrinsically forced extractions corresponding to first events of newly entering explorers. The fraction of such non-uniform selections scales as
\[
    \phi(t) = f(D)\,\frac{dD}{dt}.
\]
For non-explosive branching processes and for the sublinear or marginal novelty growth regimes considered here, one has $\phi(t)\to 0$ asymptotically. Hence this correction vanishes in intrinsic time, and the dynamics becomes effectively equivalent to uniform sampling among the active pool of explorers. If this term did not vanish asymptotically, a finite fraction of extractions would systematically correspond to first events of newly introduced explorers. However, by construction the cumulative number of distinct explorers cannot exceed the total number of intrinsic extractions, i.e. $D_w(t) \leq t$.

The microscopic dynamics is then summarized by three coupled equations:
\[
        \frac{dD}{dt} = \frac{\nu D}{\rho t + \nu D}, 
        \qquad 
        \frac{dw}{dD} = f(D),     
        \qquad 
        \frac{dt}{d\tau} = a\, w(t(\tau)),
\]
describing, respectively, the UMT discovery rate, the growth of the explorer population, and the mapping between intrinsic and natural time. In intrinsic time, the UMT admits the standard Heaps' law solution:
\[
            D(t) \sim t^{\beta}, \qquad \beta = \nu/\rho,
\]
which depends solely on microscopic urn parameters and is unaffected by the branching dynamics. To express the dynamics in real time, we use:
\[
        \frac{dD}{d\tau} = 
        \frac{dD}{dt}\,\frac{dt}{d\tau}.
\]
From $D(t)\sim t^{\beta}$ we obtain $t \sim D^{\gamma}$ with $\gamma = 1/\beta$, so that:
\[
        \frac{dD}{dt} \sim \frac{1}{\gamma} D^{1-\gamma}.
\]
Substituting into the expression above gives:
\begin{equation}
        \frac{dD}{d\tau}
            = \frac{a}{\gamma}\, D^{\,1-\gamma}\, w(\tau),
        \label{eq:dDdtau_final}
\end{equation}
showing that macroscopic growth in natural time is governed by the combined effect of the microscopic scaling exponent $\gamma$ and the branching dynamics of the explorer population $w(\tau)$. Different functional forms of $w(D)$ therefore lead to qualitatively distinct macroscopic regimes.

We analyze three representative cases---logarithmic, linear, and exponential branching---each producing a characteristic macroscopic regime, as illustrated in Fig.~\ref{fig:fig5}. Throughout this section we use the conventions $a = 1$ and $w_0 = 0$, so that $w(\tau)$ corresponds directly to the dynamically generated explorer population $D_w(\tau)$.

\subsubsection*{Logarithmic branching}

A first non-trivial regime is obtained when the explorer population grows logarithmically with the number of novelties:
\[
            \frac{dw}{dD} = \frac{1}{D}
            \qquad\Longrightarrow\qquad
            w(D)=\log D.
\]
Equ.~\eqref{eq:dDdtau_final} becomes:
\[
            \frac{dD}{d\tau}=\frac{1}{\gamma}\,D^{\,1-\gamma}\,\log D.
\]
For the microscopic regime relevant here ($\beta<1$, hence $\gamma>1$), the logarithmic factor is asymptotically negligible:
\[
            \frac{dD}{d\tau}\sim \frac{\log D}{D^{\,\gamma-1}}\to 0\quad (D\to\infty),
\]
so the dynamics in natural time remains sublinear and effectively identical to the intrinsic Heaps-like scaling: $D(\tau)\sim \tau^{\beta}$. A marginally superlinear case occurs when $\beta=1$ ($\gamma=1$), producing:
\[
            \frac{dD}{d\tau}=\log D,
            \qquad 
            Li(D)\sim\tau,
\]
where $Li$ is the logarithmic integral.

\subsubsection*{Linear branching}

A simple and illustrative case is the linear growth of explorers:
\[
            \frac{dw}{dD}=b,
\]
meaning that, on average, $b$ new explorers enter the system per novelty. This gives $w(D)\sim D$, and Eq.~\eqref{eq:dDdtau_final} becomes:
\[
            \frac{dD}{d\tau}=\frac{1}{\gamma}\,D^{\,2-\gamma}.
\]
If $\gamma=1$ (microscopic exponent $\beta=1$), the solution is exponential:
\[
            D(\tau)\sim e^{\tau},
\]
while for $\gamma>1$ the system displays superlinear but sub-exponential growth, with an exponent controlled by $2-\gamma$. Linear branching, therefore, represents the minimal mechanism capable of producing macroscopic accelerations from purely sublinear microscopic exploration.

\subsubsection*{Exponential branching}
\begin{figure*}[ht!]
    \centering
    \includegraphics[width=\linewidth]{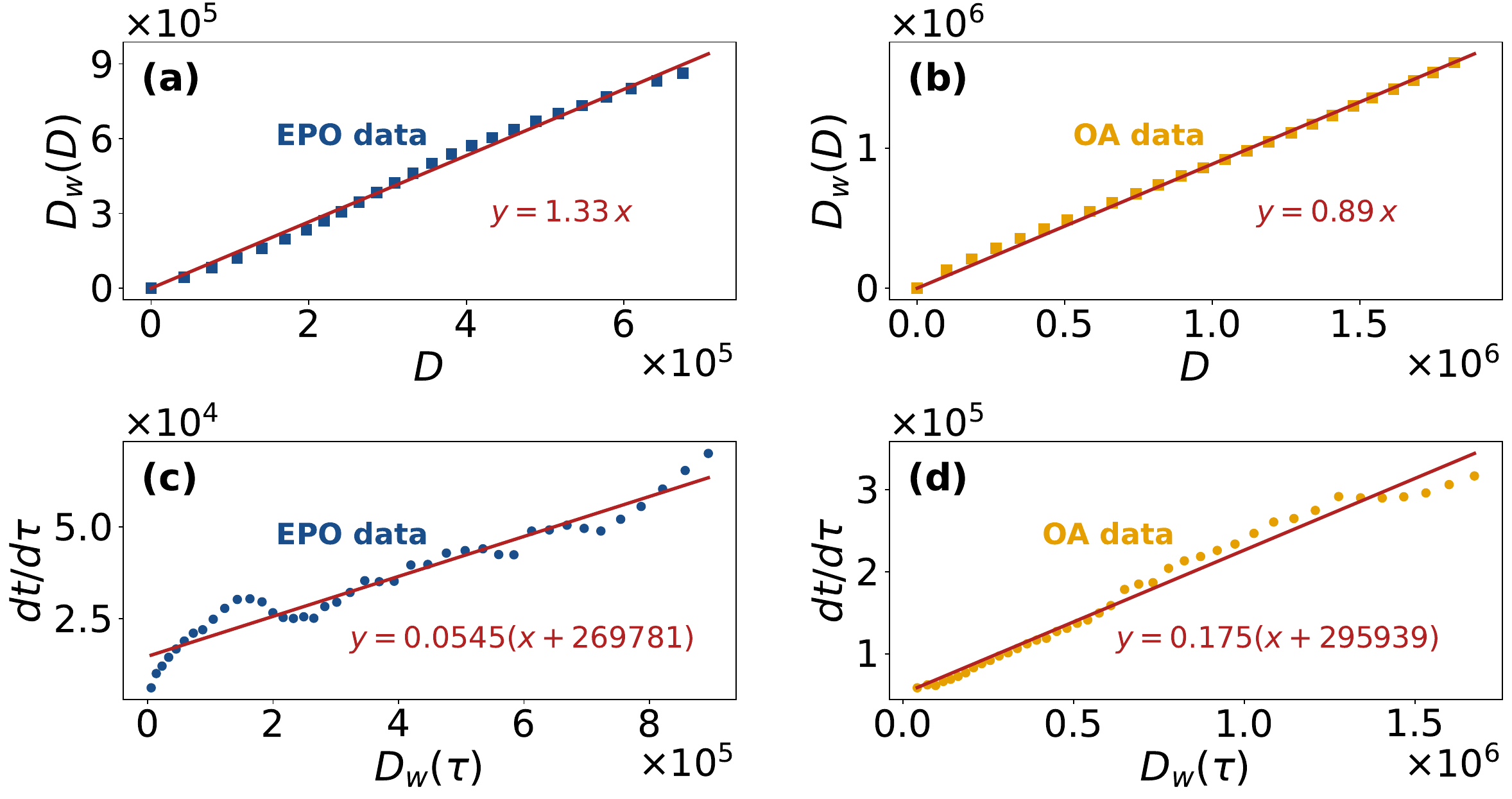}
    \caption{
    \textbf{Calibration of branching and productivity parameters.} Panels \textbf{(a–b)} show the empirical relationship between the number of explorers $D_w$ and the number of novelties $D$, used to calibrate the branching process. The predicted linear dependence $D_w = bD$ fits the data well for both patents and papers, giving $b = 1.33$ and $b = 0.89$, respectively, with all $p$-values below $10^{-10}$.  Panels \textbf{(c--d)} calibrate the productivity relation $dt/d\tau = a\,(w_0 + D_w(\tau))$ by regressing the yearly event count $\delta t(\tau)$ against the number of active explorers. This yields $a = 0.0545$, $w_0 \approx 2.70\times10^{5}$ for patents and $a = 0.175$, $w_0 \approx 2.96\times10^{5}$ for papers, again with $p$-values below $10^{-10}$. The large $w_0$ values reflect the substantial pre-existing explorer populations at the start of the observation window.}
\label{fig:fig6}
\end{figure*}
When the explorer population grows proportionally to its current size:
\[
            \frac{dw}{dD}\sim w,
            \qquad\Longrightarrow\qquad
            w(D)\sim e^{D},
\]
the macroscopic dynamics changes qualitatively. Eq.~\eqref{eq:dDdtau_final} becomes:
\[
            \frac{dD}{d\tau}
                =\frac{1}{\gamma}\,D^{\,1-\gamma}\,e^{D}.
\]
Integration gives the implicit solution:
\[
            \Gamma(\gamma-1,\,D)\sim\tau,
\]
with $\Gamma$ the incomplete gamma function. Since $\Gamma(\gamma-1,D)$ has a vertical asymptote, $D(\tau)$ necessarily diverges at a finite time. For the special case $\gamma=1$:
\[
            \frac{dD}{d\tau}=e^{D}
            \quad\Rightarrow\quad
            D(\tau)\sim \log\!\left(\frac{1}{1-\tau}\right),
\]
which diverges as $\tau\to 1$.  
            
This regime generates the typical finite-time singularities of TAP-like dynamics~\cite{cortes2022biocosmology,cortes2025tap, korotayev202021st, kurzweil2005singularity}: explosive growth arises directly from the rapid proliferation of explorers, producing a nonlinear mapping between intrinsic and natural time and driving macroscopic innovation accelerations.

\subsection{Calibration of the model}
\label{sec:methods_calibration}

The model parameters governing the growth of the explorer population and the productivity relation are estimated directly from the empirical trajectories (Fig.~\ref{fig:fig6}).  
        
We first determine the branching parameter $b$ from the empirical proportionality between the number of distinct explorers and the number of discovered elements, $D_{w} \sim b D$. A regression constrained through the origin yields:
\[
        b = 1.33 \quad\text{(patents)}, \qquad 
        b = 0.89 \quad\text{(papers)},
\]
with $p$-values below $10^{-10}$ in both cases, confirming the robustness of the linear branching assumption.
        
We then calibrate the productivity relation:
\[
        \frac{dt}{d\tau} = a\,(w_0 + D_w(\tau))
\]
by regressing the yearly event count $\delta t(\tau)$ against $D_w(\tau)$.  The slope provides $a$, while the intercept gives $a w_0$. This gives:
\[
        a = 0.0545, \qquad w_0 = 269781
        \quad\text{(patents)},
\]
\[
        a = 0.175, \qquad w_0 = 295939 
        \quad\text{(papers)},
\]
again with $p$-values below $10^{-10}$. The large values of $w_0$ reflect the fact that both technological and scientific systems were already highly populated at the start of the observation window. 
        
Together with the microscopic urn parameters (set to $\nu=\rho=1$), these estimates fully specify the model for comparison with the empirical trajectories.

\subsection{Initial conditions}
\label{sec:methods_initial_cond}

Because both patenting and scientific publishing were already mature systems by the start of our observation window (1980), the early part of the empirical trajectories displays a transient phase in which novelties accumulate faster than in the asymptotic urn dynamics. To account for this, we initialize the model with a large effective number of pre-existing elements, $N_{0}$, representing the unobserved innovation accumulated before 1980.
        
To preserve the heterogeneous composition of the innovation space, the initial state cannot be assumed uniform. We generate it by running the same microscopic urn dynamics up to $D(t)=N_{0}$ and using the resulting element frequency distribution as the starting condition at $\tau=0$. Elements that were already common before 1980 thus begin with higher multiplicity. Formally, we assign each element an initial weight $N_{\rho}$ and slightly generalize the UMT update rule: when such an element is discovered, all its $N_{\rho}$ balls are added at once, and the adjacent possible is expanded by exactly $\nu$ new elements. In this formulation, $\nu$ retains its interpretation as the net growth rate of the adjacent possible, independent of the initial weight distribution. Matching the empirical transient gives:
\[
        N_{0}\approx 1.2\times 10^{5}\ \text{(patents)}, \qquad
        N_{0}\approx 1.3\times 10^{4}\ \text{(papers)}.
\]
These initial conditions are used for all simulations and analytical comparisons.

% \bibliography{bibliography}

% \section*{Acknowledgements}

\section*{Author contributions statement}

V.T. conceived the original idea and supervised the study. A.B. and G.D.M. developed the initial analytical and numerical framework. G.D.B. and G.D.M. provided the datasets and contributed to the modeling. A.B. and G.D.B. performed the simulations and empirical analyses and prepared the first draft of the manuscript. All authors discussed the results and contributed to revising and improving the manuscript and figures.

\section*{Additional information}

The code used for the simulations and for the empirical data analyses is publicly available at: \url{https://github.com/alebellina412/modelling_micro_macro}.

\clearpage
\newpage

%%%%%%%%%%%%%%%%%%%%%%%%%%%%%%%%%%%%%%
%%%%%%% SUPPLEMENTARY MATERIAL %%%%%%%
%%%%%%%%%%%%%%%%%%%%%%%%%%%%%%%%%%%%%%

\setcounter{figure}{0}
\setcounter{table}{0}
\setcounter{equation}{0}
\setcounter{section}{0}
\makeatletter
\renewcommand{\thefigure}{S\arabic{figure}}
\renewcommand{\theequation}{S\arabic{equation}}
\renewcommand{\thetable}{S\arabic{table}}

\renewcommand{\thesection}{S\arabic{section}}

\setcounter{secnumdepth}{2} 
\onecolumn

% \widetext
\begin{center}
\textbf{\Large Supplementary Information for\\"The gold-rush effect: how innovation speeds up"}
\end{center}

\section{Inter-event times and entry--exit dynamics}

In the baseline version of the model in this paper, $a w(\tau)$ explorers (authors or inventors) at each intrinsic-time step are sampled from the pool of available explorers $w(t)$. The extractions are uniformly random, except for the branching-induced activations, which introduce a small fraction of forced extractions quantified by $\phi(t)=f(D)\,dD/dt$. In the case considered in this manuscript, $f(D)=b$ (linear branching), while the exact expression for the number of novelties in intrinsic time is $D(t)\sim t/\log t$, giving $dD/dt\sim 1/\log t$~\cite{tria2014dynamics}. Hence, $\phi(t)=b/\log t$, which vanishes for large intrinsic time.

The effective per-explorer event probability is therefore $a(1-\phi(t))$, implying an average inter-event time
\[
\langle \ell \rangle 
= \frac{1}{a(1-\phi(t))}
\simeq \frac{1}{a}\left(1+\frac{b}{\log t}\right)
\longrightarrow \frac{1}{a}.
\]
Thus, to leading order in intrinsic time, the average inter-event time is $\langle \ell \rangle \simeq 1/a$ (inverse of velocity) for an author chosen uniformly from the population. While this assumption is sufficient to reproduce the macroscopic growth of novelties and the average productivity per active explorer over the considered time span, it leads to unrealistically large inter-event times at the individual level when compared with empirical data. Specifically, the baseline model predicts $\langle \ell \rangle \simeq 18.35$ years for patents and $\langle \ell \rangle \simeq 5.71$ years for papers, to be contrasted with the empirical values $2.34$ and $1.53$ years, respectively.

Empirically, individual activity is temporally correlated: publications and patents are typically produced within finite careers, characterized by relatively short inter-event times during active periods and permanent exit afterwards. To account for this discrepancy while keeping the microscopic urn dynamics and the macroscopic growth laws unchanged, we introduce a minimal entry--exit dynamics.

Specifically, we introduce an exit probability $\mu$: at each time step $\tau$, an explorer exits the system with probability $\mu$. This naturally distinguishes between the total number of explorers ever observed, $w(\tau)$, and the number of explorers still active (i.e., not exited), $w_{\mathrm{alive}}(\tau)$. The dynamics of $w_{\mathrm{alive}}(\tau)$ reads
\begin{equation}
    w_{\mathrm{alive}}(\tau+1) = (1-\mu)\, w_{\mathrm{alive}}(\tau) + B(\tau+1),
    \qquad
    \dot w_{\mathrm{alive}}(\tau) = -\mu\, w_{\mathrm{alive}}(\tau) + B(\tau),
    \label{eq:w_alive}
\end{equation}
where $B(\tau)$ is the branching contribution defined in the main text. The total number of explorers evolves as usual according to
\[
    w(\tau+1) = w(\tau) + B(\tau+1),
    \qquad
    \dot w(\tau) = B(\tau).
\]

The presence of an exit dynamics modifies the average inter-event time at the individual level, which can be written as
\[
    \ell(\tau) = \frac{w_{\mathrm{alive}}(\tau)}{w_{\mathrm{active}}(\tau)}
            = \frac{w_{\mathrm{alive}}(\tau)}{a\, w(\tau)} .
\]
In the case of the simple branching process, using $w(\tau) = w_0 e^{ab\tau}$ and $B(\tau) = ab\, w_0 e^{ab\tau}$, solving Eq.~\eqref{eq:w_alive} yields
\[
    w_{\mathrm{alive}}(\tau)
    = w_0\, \frac{ab\, e^{ab\tau} + \mu\, e^{-\mu\tau}}{ab+\mu}.
\]
For $\tau \gg 1$, the transient term can be neglected, leading to
\[
    w_{\mathrm{alive}}(\tau) \simeq \frac{ab}{ab+\mu}\, w(\tau),
    \qquad
    \ell \equiv \lim_{\tau\to\infty} \ell(\tau) = \frac{b}{ab+\mu}.
\]
This relation allows us to infer the exit probability $\mu$ by matching the empirical inter-event time $\ell$, giving
\[
    \mu = \frac{b}{\ell} - ab,
    \qquad
    \mu_{\mathrm{patents}} = 0.50,
    \qquad
    \mu_{\mathrm{papers}} = 0.43,
\]
which automatically reproduces the empirical values $\ell_{\mathrm{patents}}=2.34$ and $\ell_{\mathrm{papers}}=1.53$. 

We can now compute the average career length for explorers. In doing so, we restrict both the empirical analysis and the model-based estimate to explorers that have produced at least two events, for which the career length is well defined and equals $L_i = \tau_i^{\mathrm{last}} - \tau_i^{\mathrm{first}}$, i.e., the difference between the last and first event. The entry--exit dynamics naturally generates a large fraction of one-timers, i.e., explorers producing exactly one event. After the first event, an explorer exits before producing a second one with probability
\[
    f \equiv \mathbb{P}(n=1)
    = \frac{\mu}{\mu + (1-\mu)p},
    \qquad p = \frac{1}{\ell},
\]
which defines the fraction of one-timers. Using the parameters above, we obtain $f_{\mathrm{patents}} = 0.70$ and $f_{\mathrm{papers}} = 0.53$, in good agreement with the empirical values $\bar f_{\mathrm{patents}}=0.75$ and $\bar f_{\mathrm{papers}}=0.55$.

Finally, we estimate the typical career span of explorers with more than one event in the model. For each explorer $i$, the career length can be approximated as $L_i = n_i \ell_i$, i.e., the number of events times the average inter-event time. In the model, explorers are sampled uniformly at random and inter-event times are homogeneous on average, so that $\ell_i \simeq \ell$ and thus $L_i \simeq n_i \ell$. Taking the average over explorers with at least two events, we obtain
\[
    L_{\mathrm{model}} \simeq \langle n_i \rangle \ell = \frac{N_2}{w_2} \ell,
\]
where $N_2$ is the total number of events produced by explorers with at least two events, and $w_2$ is the number of such explorers. These quantities can be computed as $N_2 = t(\tau=40) - f w(\tau=40)$ and $w_2 = (1-f) w(\tau=40)$, giving the estimates $L_{\mathrm{model}}^{\mathrm{patents}} = 3.82$ and $L_{\mathrm{model}}^{\mathrm{papers}} = 9.86$ (using the values from the simulation: $t(\tau=40)=1,384,928$ and $D_w(\tau=40)=892,023$ for patents, and $t(\tau=40)=6,617,495$ and $D_w(\tau=40)=1,568,791$ for papers). These values are compared with the empirical estimates from the data $L_{\mathrm{data}} = \langle \tau_i^{\mathrm{last}} - \tau_i^{\mathrm{first}} \rangle$, giving $L_{\mathrm{data}}^{\mathrm{patents}} = 5.13$ and $L_{\mathrm{data}}^{\mathrm{papers}} = 9.36$, in good accordance. This comparison shows that the introduction of the entry--exit mechanism restores career lengths to a realistic finite scale, in contrast with the baseline model, where individual careers are effectively unbounded.

\section{Heterogeneous individual exploration rates}
\begin{figure*}[t!]
    \centering
    \includegraphics[width=\linewidth]{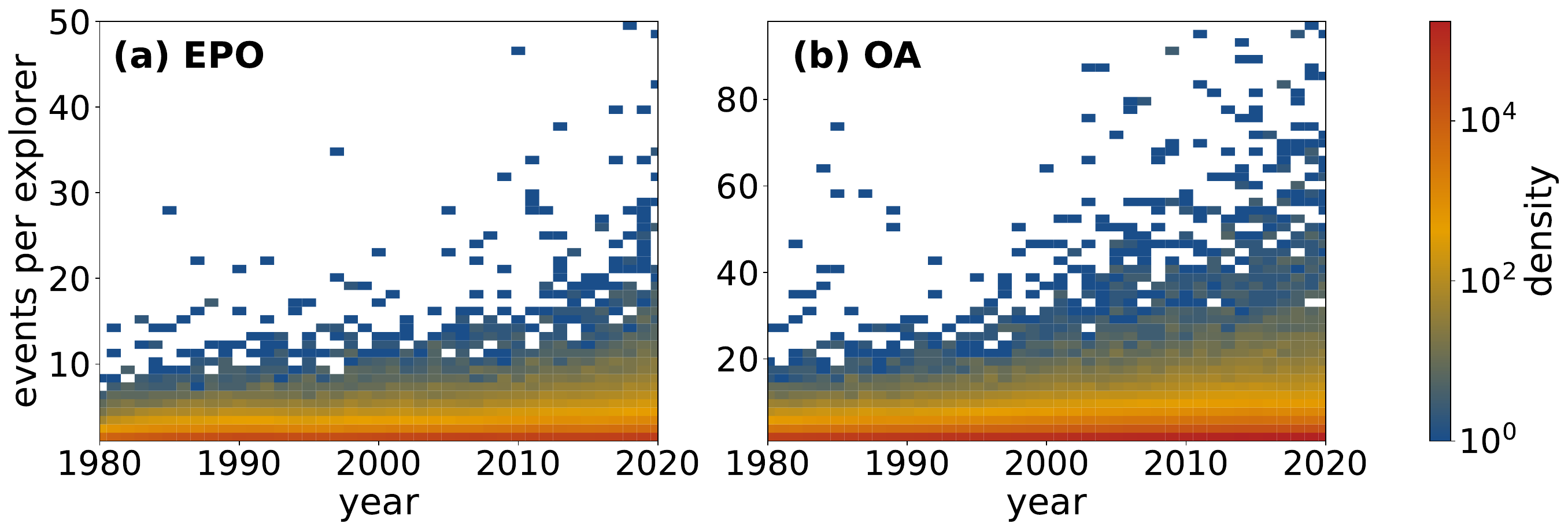}
    \caption{\textbf{Heterogeneity of individual yearly production.} Density of the number of events produced per explorer per year in the EPO patent dataset (left) and in the OpenAlex publication dataset (right). While the distributions are strongly concentrated around one event per year, indicating an approximately constant average productivity, they display a broad dispersion across individuals, with the presence of highly productive outliers.}
\label{fig:figSI1}
\end{figure*}

In the model, explorers are described as statistically uniform agents. Each active explorer generates events according to the same stochastic dynamics, characterized by a typical inter-event time, so that individual trajectories differ only because of stochastic fluctuations and entry or exit times in the extended version of the model proposed in the previous section. This modeling choice provides a minimal and parsimonious description of microscopic exploration.

Empirical data reveal much broader variability at the individual level. Figure~\ref{fig:figSI1} shows the distribution of the number of events produced per year by individual explorers in scientific and technological datasets. While the density is strongly concentrated around one event per year (consistent with the fact that the average productivity per active explorer is close to unity, as shown in Fig.~\ref{fig:fig1}), there exists a wide dispersion across individuals, with a non-negligible fraction of highly productive authors and inventors. This indicates that, in real systems, explorers follow heterogeneous microscopic dynamics, with different effective exploration rates.

A more detailed modeling framework could explicitly account for this heterogeneity by assigning individual exploration velocities or by drawing inter-event times from a broad distribution, possibly calibrated on empirical data. Such an extension would allow the model to reproduce the full distribution of individual outputs and the presence of extreme productivity values observed in real careers.

In the present work, we deliberately adopt a simplified description in which explorers are statistically uniform. This choice is motivated by the focus on collective dynamics: macroscopic observables such as the total number of events $t(\tau)$, the number of active explorers $w(\tau)$, and the number of distinct discoveries $D(\tau)$ depend only on the aggregate number of events occurring in natural time, and not on how these events are distributed across individuals. Accordingly, we stress that introducing heterogeneity in individual exploration rates, or in entry--exit dynamics such as career termination, would primarily affect microscopic trajectories, while leaving the collective phenomenology unchanged.

\end{document}